\documentclass[12pt,a4,sans]{article}
\usepackage[latin1]{inputenc}
\usepackage{amssymb}
\usepackage{amsmath}
\usepackage{amsthm}
\usepackage{amsfonts}
\usepackage[pdftex]{graphicx}
\usepackage{geometry}
\usepackage{braket}
\usepackage{enumerate}
\usepackage[usenames,dvipsnames]{xcolor}
\usepackage{setspace}
\usepackage[backgroundcolor=white,linecolor=black]{todonotes}
\usepackage[in]{fullpage}
\usepackage{hyperref}
\usepackage{array}
\usepackage{cancel}
\usepackage{wrapfig}
\usepackage[font=small,labelfont=bf]{caption}

\numberwithin{equation}{section}

\hypersetup{
	colorlinks=true,
	linktoc=page,
	citecolor=Blue,
	linkcolor=Blue,
	urlcolor=Blue} 

\makeatletter 
\renewcommand{\maketitle} 
{ \begingroup \begin{center} \large {\bf \@title}
		\vskip 5pt \large \@author \\ \vskip 5pt \@date \end{center}
	\vskip 5pt \endgroup \setcounter{footnote}{0} }
\makeatother 

\newcommand{\comments}[1]{}

\newcommand{\ra}{\rangle}

\newcommand{\N}{\mathcal{N}}

\newcommand{\Tr}{\text{Tr}}

\renewcommand{\b}[1]{\braket{#1}}

\renewcommand{\O}{\mathcal{O}}

\newcommand{\be}{\begin{equation}}
\newcommand{\ee}{\end{equation}}

\def\beqa{\begin{eqnarray}}
\def\eeqa{\end{eqnarray}}
\def\beq{\begin{equation}}
\def\eeq{\end{equation}}

\def\Tr{{\rm Tr}}
\def\one{\mbox{1 \kern-.59em {\rm l}}}

  \def\cL{{\cal L}}
 \def\cN{{\cal N}} \def\cO{{\cal O}}
  \def\cR{{\cal R}}
 \def\cT{{\cal T}}

\def\uno{\mbox{1 \kern-.59em {\rm l}}}

\def\one{1\!\!1\,\,}

\def\bcomment#1{}

\def\eps{\epsilon}

\usepackage{slashed}
\usepackage{caption}

\usepackage[nottoc,notlot,notlof]{tocbibind}
\usepackage[nosort]{cite}
\usepackage{color}

\usepackage{parskip}

\graphicspath{{./Images/}}

\long\def\symbolfootnote[#1]#2{\begingroup%
	\def\thefootnote{\fnsymbol{footnote}}\footnote[#1]{#2}\endgroup}

\begin{document}
	
	\begin{flushright}
		QMUL-PH-17-12\\
		CERN-TH-2017-155\\
		HU-EP-17/18\\
		NSF-ITP-17-090
	\end{flushright}
	
	\vspace{15pt} 

	\begin{center}
		
		{\Large \bf    Higgs amplitudes from $\mathcal{N}\!=\!4$ super Yang-Mills theory}\\

		%
		\vspace{25pt}

		
		{\mbox {\bf  Andreas~Brandhuber$^{a,b,\S}$, 
				Martyna~Kostaci\'{n}ska$^{a,\S}$,}} \\ \vspace{0.2cm}
				{\mbox{\bf
				Brenda~Penante$^{c,\star}$ and  %
				Gabriele~Travaglini$^{a,b,d,\S}$}}%

		\vspace{0.5cm}
		
		\begin{quote}
			{\small \em
				\begin{itemize}
					\item[\ \ \ \ \ \ $^a$]
					\begin{flushleft}
						Centre for Research in String Theory\\
						School of Physics and Astronomy\\
						Queen Mary University of London\\
						Mile End Road, London E1 4NS, United Kingdom
					\end{flushleft}
					
					\item[\ \ \ \ \ \ $^b$]Kavli Institute for Theoretical Physics\\ University of California, Santa Barbara, CA 93106, USA
					
					\item[\ \ \ \ \ \ $^c$]CERN Theory Division, 1211 Geneva 23, Switzerland
					
					\item[\ \ \ \ \ \ $^d$]
					Institut f\"{u}r Physik und IRIS Adlershof\\
					Humboldt-Universit\"{a}t zu Berlin\\
					Zum Gro{\ss}en Windkanal 6, 12489 Berlin, Germany

				\end{itemize}
			}
		\end{quote}


		\vspace{15pt}  

{\bf Abstract}
	\end{center}
	
	\vspace{0.3cm} 
	
\noindent

\noindent
Higgs plus multi-gluon amplitudes in QCD can be computed in an effective Lagrangian description. 
In the infinite top-mass limit,  an amplitude with a Higgs and $n$ gluons  is computed by  the form factor of the operator ${\rm Tr}\, F^2$. Up to  two loops and for three gluons, its maximally transcendental part  is captured entirely by the form factor of the protected stress tensor multiplet operator $\mathcal{T}_2$ in $\cN\!=\!4$ supersymmetric Yang-Mills theory. 
The next order correction involves the calculation of the form factor of the higher-dimensional, trilinear operator  ${\rm Tr}\, F^3$. We present explicit results at two loops for three gluons, including the subleading transcendental terms derived from a particular descendant of the Konishi operator that contains ${\rm Tr}\, F^3$. These are expressed in terms of a few universal building blocks already identified in earlier calculations. 
We show   that the maximally transcendental part of this quantity, computed in non-supersymmetric Yang-Mills theory, is identical to   the form factor of another  protected operator, $\mathcal{T}_3$, in the maximally supersymmetric theory. 
Our results  suggest that the maximally transcendental  part of Higgs amplitudes in QCD can be entirely computed through $\cN\!=\!4$ super Yang-Mills.

	\vfill
	\hrulefill
	\newline
\vspace{-1cm}
$^{\S}$~\!\!{\tt\footnotesize\{a.brandhuber, m.m.kostacinska, g.travaglini\}@qmul.ac.uk}, \ $^{\star}$~\!\!{\tt\footnotesize b.penante@cern.ch}

	\setcounter{page}{0}
	\thispagestyle{empty}
	\newpage


	


\section{Introduction}\label{sec:Introduction}
	
	Whether or not it will be discovered in present or future searches at the Large Hadron Collider (LHC), supersymmetry is a powerful organisational principle of perturbative calculations in quantum field theory.
	One example of such success is the  one-loop supersymmetric decomposition \cite{Bern:1994cg}, whereby the calculation of a one-loop scattering amplitude in pure Yang-Mills theory --  a crucial ingredient for  constructing QCD amplitudes --  is traded for three simpler calculations: that of the same amplitude in the 
$\cN\!=\!4$ (or maximally) supersymmetric Yang-Mills (SYM)  theory, 
	plus the contributions of an $\cN\!=\!1$ chiral multiplet and a scalar running in the loop. The technical difficulty in dealing with gluons in the loop is  thus replaced by three simpler calculations, two of which are performed in supersymmetric theories. 

	Supersymmetry makes a remarkable   appearance in the principle of maximal transcendentality \cite{Kotikov:2002ab,Kotikov:2004er}, allowing anomalous dimensions of twist-two operators in $\cN\!=\!4$ SYM to be obtained from those computed in QCD \cite{Moch:2004pa,Vogt:2004mw}
	by simply deleting all terms with degree of transcendentality less than  maximal  ($2L$ at $L$ loops).  
	Conversely, one can say that $\cN\!=\!4$ SYM captures the ``most complicated", or  maximally transcendental part of this QCD result. 
	Alas, scattering amplitudes in general do not  satisfy the principle of maximal transcendentality. For instance, an $n$-point MHV amplitude computed in pure Yang-Mills for generic $n$  receives additional contributions that have maximal transcendental degree   already at one loop  \cite{Bern:1994cg, Bern:1993mq,Bedford:2004nh}.  
	
	Multi-gluon Higgs amplitudes seem to provide a fortunate exception    where the   principle of maximal transcendentality may in fact apply \cite{Brandhuber:2012vm}. 
	To discuss this, we recall that gluon fusion  through a top-quark loop is the dominant mechanism for Higgs production at the LHC; in  an approximation where the mass of the top,  $m_t$,  is  much larger than the mass of the Higgs, $m_H$, an effective Lagrangian description can be used to compute such amplitudes. The leading order term is a dimension-five operator $\mathcal{L}^{(0)} \sim H\, {\rm Tr} \,F^2$, where $H$ represents the  Higgs field and $F$ is the gluon field strength
	\cite{Wilczek:1977zn,Shifman:1979eb,Dawson:1990zj}.  Hence,  Higgs plus multi-gluon amplitudes at leading order are form factors of ${\rm Tr}\, F^2$.   
	The surprising result of \cite{Brandhuber:2012vm} is   that the form factor of an ``appropriate  translation" of the operator ${\rm Tr}\, F^2$ to $\cN\!=\!4$ SYM, computed  in the maximally supersymmetric  theory, 
is identical to the maximally transcendental part\footnote{The same maximally transcendental part appears in a non-minimal two-loop form factor of the Konishi operator in $\cN\!=\!4$ SYM \cite{Banerjee:2016kri}.}
of the Higgs plus three gluon amplitude in QCD of  \cite{Gehrmann:2011aa}, and is independent of the gluon helicities.
This appropriate translation turns out to be the simplest composite operator in the theory, namely the stress tensor multiplet operator $\mathcal{T}_2$. It is  protected from quantum corrections (or half-BPS), and as such it does not mix with any other operator.  Furthermore, its form factors have only infrared divergences. Two components of $\mathcal{T}_2$ are particularly relevant here:  the chiral on-shell Lagrangian\footnote{The precise expression of $\cL_\text{on-shell}$  can be found in  \cite{Eden:2011yp,Brandhuber:2011tv}.}, $\cL_{\text{on-shell}} \ni {\rm Tr}F^{2}_{\rm ASD}$, and ${\rm Tr}\,X^2$, where    $F_{\rm ASD}$ is the anti-selfdual part of the gluon field strength and $X$ is one of the three complex scalar fields of $\cN\!=\!4$ SYM. Note that it is $\cL_\text{on-shell}$ that does not mix under renormalisation, and not~${\rm Tr}\,F^{2}_{\rm ASD}$. 

In parallel,  one can  study    
subleading corrections in $m_H/m_t$, which  in the effective Lagrangian setup are  captured by higher-dimensional operators. The first corrections arise at dimension seven and include the interactions $\mathcal{L}^{(1)} \sim H\, {\rm Tr} \,F^3$ and 
$\mathcal{L}^{(2)} \sim H\,  {\rm Tr} (D_\mu F_{\nu \rho}D^\mu F^{\nu \rho})$ 
\cite{Buchmuller:1985jz, Neill:2009tn, Neill:2009mz, Harlander:2013oja, Dawson:2014ora}. 
In this paper
 we compute the two-loop form factor of an appropriate translation    to the $\cN\!=\!4$ theory of the operator ${\rm Tr}\, F^3$ in the case of three gluons. Our key finding is that its
maximally transcendental part is
identical to that of the  contribution 
arising from the operator $\mathcal{L}^{(1)}\!\sim\!H\, {\rm Tr} \,F^3$ 
to the Higgs plus multi-gluon amplitude in QCD.  As we shall see, this maximally transcendental contribution    turns out to be  identical to the form factor of another special  operator, namely  the trilinear half-BPS operators $\mathcal{T}_3$. This is  an appropriate supersymmetrisation of ${\rm Tr} \, X^3$, whose minimal form factor has been  computed in  \cite{Brandhuber:2014ica} at two loops.
  Hence the simplest   operators in the maximally supersymmetric  theory in four dimensions,  the half-BPS operators, compute the most complicated, or maximally transcendental part of  the non-supersymmetric Higgs plus multi-gluon amplitudes. 

To identify this appropriate translation, we observe that  ${\rm Tr}\, F_{\rm ASD}^3$ is a trilinear, non-protected operator which  at one loop has the same anomalous dimension as the  Konishi operator.  
A natural choice is to take the descendant  obtained by acting with eight $\bar{Q}$-supersymmetries on the Konishi operator $\epsilon^{ABCD} {\rm Tr}(\phi_{AB} \phi_{CD})$, landing on ${\rm Tr} \, F_{\rm ASD}^3$
dressed with appropriate  additional terms as required by  supersymmetry. Here $\phi_{AB}$  denote the scalar fields of the theory, with $A, \ldots, D = 1, \ldots , 4$ being fundamental $SU(4)$ indices. 
Note that since this descendant is obtained by acting with tree-level supersymmetry generators,  any potential mixing   is deferred to one  loop. We will also pick an external state containing three positive-helicity gluons -- a state that is produced by ${\rm Tr} \, F_{\rm ASD}^3$ acting on the vacuum.  Tree-level form factors of the full Konishi multiplet  have recently been studied in \cite{Koster:2016loo,Chicherin:2016qsf}.

 An earlier two-loop calculation is also relevant here: in \cite{Brandhuber:2016fni} we considered the form factors of a particular trilinear descendant of the Konishi operator made mostly of scalar fields, rather than field strengths, namely $\cO_{\mathcal{K}} = \mathcal{O}_B - gN/ (8 \pi^2)\mathcal{O}_F$, where 
$\mathcal{O}_B := {\rm Tr} (X[Y , Z])$ and  $\mathcal{O}_F := (1/2) {\rm Tr} (\psi \psi)$. 
The three scalar fields $X:=\phi_{12}$,  $Y:=\phi_{23}$, $Z:=\phi_{31}$ and the fermion $\psi_\alpha := \psi_{123, \alpha}$ are the letters of the $SU(2|3)$ closed subsector of the $\cN\!=\!4$ theory \cite{Beisert:2003ys} 
(the second  term in $\cO_{\mathcal{K}}$ is induced by mixing, and does not contribute to the maximally transcendental part of the result).
The maximally transcendental part of the form factor of   $\cO_{\mathcal{K}}$ is identical to that of $\cT_3$. It is accompanied by additional terms that are subleading in transcendentality, which feature in our discussion below. 

Our results can be summarised as follows.
First, the infrared-finite two-loop remainder of the form factor of the Konishi descendant containing ${\rm Tr}\, F_{\rm ASD}^3$, with a state of three positive-helicity gluons,  has  maximal degree of transcendentality equal to  four. 
Its maximally transcendental part is identical to the remainder of the half-BPS operator $\cT_3$ of \cite{Brandhuber:2014ica}, and to that of $\mathcal{O}_B$. Remarkably, the universality of this contribution  was also found in 
  \cite{Loebbert:2015ova,Brandhuber:2016fni,Loebbert:2016xkw} for the three closed sectors $SU(2)$, $SU(2|3)$ and $SL(2)$ in the $\cN\!=\!4$ theory, respectively. 
  Second,  our form factor  remainder  also  contains terms of transcendentality ranging from three to zero,  similarly to the form factor of  $\mathcal{O}_B$  \cite{Brandhuber:2016fni}. Unlike the case of $\mathcal{O}_B$, our present result is  accompanied by polylogarithms  multiplied by ratios of kinematic invariants. 
 We find that only a few universal building blocks are needed to describe all such contributions and,  interestingly, they are the same as those which appeared in   \cite{Brandhuber:2016fni} as well as in related computations of the spin chain Hamiltonian performed in 
 \cite{Loebbert:2015ova,Loebbert:2016xkw}, suggesting the universality of these quantities. Finally, we observe that the computation of the four-dimensional cut-constructible part of the form factor of ${\rm Tr} \, F^3$ in QCD differs from our calculation 	in $\cN\!=\!4$ SYM only by certain single-scale integrals  of sub-maximal transcendentality, in the three-gluon case considered in detail here. Hence $\cN\!=\!4$ SYM captures the 
 maximally transcendental part not only of the leading-order Higgs plus three-gluon amplitudes, as found in \cite{Brandhuber:2012vm}, but also of the subleading corrections arising from ${\rm Tr} \, F^3$. That the maximally supersymmetric theory may be relevant for computing phenomenologically interesting amplitudes is a happily surprising result.

 The rest of the paper is organised as follows. In Section \ref{Sec:2} we introduce the building blocks of our two-loop calculation, including the tree-level and one-loop form factors of the relevant operators, and discuss the methodology used to derive the result
(details of the calculation will appear in \cite{BKPTinprep}).
 In Section \ref{Sec:3} we present  our two-loop result. We conclude in  Section \ref{Sec:4} by discussing the modifications needed in a calculation performed in non-supersymmetric Yang-Mills and why these do not alter the maximally transcendental part of the $\cN\!=\!4$ result.

\section{Outline of the computation}
\label{Sec:2}
	
In this letter we consider form factors of the dimension six operator $\mathcal{O}_1 \sim {\rm Tr} \, F_{\rm ASD}^3 + \mathcal{O}(g)$ with three positive-helicity gluons up to two loops in $\mathcal{N}=4$ SYM. Note that ${\rm Tr} \, F_{\rm ASD}^3$ appears in the decomposition of $\Tr(F^3) = \Tr(F_{\rm SD}^3)+\Tr(F_{\rm ASD}^3)$ and the extra terms denoted by $\mathcal{O}(g)$
have length four or higher and are produced by acting with eight tree-level supercharges $\bar{Q}^A_{\dot{\alpha}}$ on the Konishi operator. In other words $\mathcal{O}_1$ is the (tree-level) descendant of the Konishi operator and any corrections due to mixing must appear at
one-loop order or higher. The overall normalisation of $\mathcal{O}_1$ is fixed so that the minimal tree-level
form factor is
\begin{align}
\label{eq:operator1}
F_{\O_1}^{(0)}(1^+,2^+,3^+;q) \, = \, [12][23][31]\, ,
\end{align}
where $q:=p_1+p_2+p_3$ is the momentum carried by the operator. For future reference we also introduce the dimensionless ratios of Mandelstam variables $u := s_{12}/q^2$, $v := s_{23}/q^2$, and $w := s_{31}/q^2$, which obey  $u+v+w=1$.
	
The other operators that can mix with $\mathcal{O}_1$ at this order and with  the particular on-shell state we have picked are ${\rm Tr}(D^\mu F^{\nu \rho}D_\mu F_{\nu\rho})$, two further operators with different Lorentz contractions, and $q^2 {\rm Tr}(F_{\rm ASD}^2)$. See \cite{Dawson:2014ora} for a discussion of suitable operator bases. In practice we need to choose a linear combination of
these operators, which we call $\mathcal{O}_2 \sim {\rm Tr} (DF DF)$, and which produces the only other possible Lorentz structure with the correct dimension and spinor weights in addition to that of  \eqref{eq:operator1}.
Explicit forms of the operators are not necessary since we use unitarity in our calculation and only tree-level form factors and amplitudes are needed as input. The relevant form factor of the appropriately normalised operator $\mathcal{O}_2$ is 
\begin{align}
\label{eq:operator2}
F_{\O_2}^{(0)}(1^+,2^+,3^+;q) \, = \, \frac{q^6}{\b{12}\b{23}\b{31}}\,  = \, - \frac{F_{\O_1}^{(0)}(1^+,2^+,3^+;q)}{u v w},
\end{align}
which is the only other possible ultraviolet  counterterm form factor.
	

The minimal one-loop form factor of the operator $\O_1$ is obtained from  two-particle cuts involving \eqref{eq:operator1} and four-point tree-level gluon amplitudes. It is given by
(see also \cite{Neill:2009mz})
\begin{align}
\label{eqn:FF-one-loop} 
\hspace{-0.6cm}F^{(1)}_{\O_1}(1^+,2^+,3^+;q)\,&=\,i\,F^{(0)}_{\O_1}(1^+,2^+,3^+;q)\Big[2\,  {\rm Bub}(s_{12})\,+\,s_{12}\, {\rm Tri}(s_{12})\,
+\,\text{cyclic}(1,2,3)\Big] \, ,
\end{align}
with 
${\rm Bub}(s) = \frac{i c_\Gamma}{\epsilon (1-2 \epsilon)} (-s/\mu^2)^{-\epsilon}$,  
${\rm Tri}(s) = \frac{i c_\Gamma}{\epsilon^2} (-s/\mu^2)^{-\epsilon}/s$, and $c_{\Gamma} = \frac{\Gamma(1+\epsilon)\Gamma^2(1-\epsilon)}{(4 \pi)^{2-\epsilon} \Gamma(1-2\epsilon)}$.
From \eqref{eqn:FF-one-loop}  we can infer the one-loop anomalous dimension $\gamma_{\O_1}^{(1)}=12\, a$, where $a = g^2 N/(4\pi)^2$ is the 't Hooft coupling.

	
We now proceed to  the minimal two-loop form factor of the operator $\O_1$. For a detailed discussion of the computation we refer the reader to the forthcoming paper \cite{BKPTinprep}.
In order to completely fix the two-loop integrand we use four types of cuts:
	\begin{figure}[h]
		\centering
		\includegraphics[width=1\textwidth]{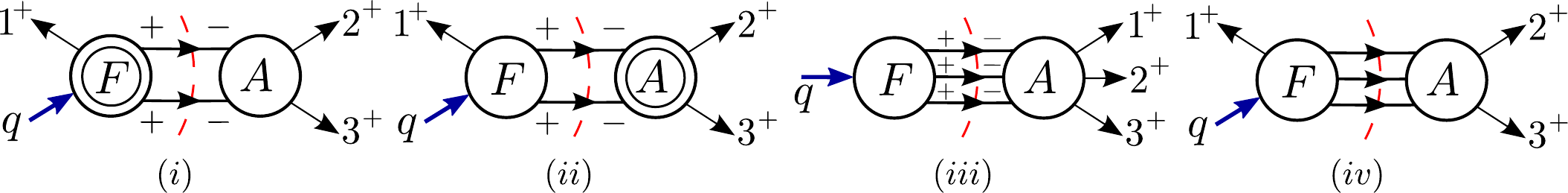}
		\caption{\it Four distinct types of cuts considered in the calculation of the two-loop form factor. In diagram $(iv)$  we sum over all possible  helicity assignments of the internal particles. We also need to include cyclic permutations of the external lines.}
		\label{fig:AllCuts}
	\end{figure}
	
	First, we consider the two-particle cut in the kinematic $s_{23}$-channel shown in Figure~\ref{fig:AllCuts}$(i)$ where as building blocks we use the one-loop form factor \eqref{eqn:FF-one-loop} and a tree-level MHV amplitude. Note that Figure~\ref{fig:AllCuts}$(ii)$ presents the two-particle cut in this channel with the tree-level form factor of \eqref{eq:operator1} and a one-loop amplitude, but this term gives no extra constraint on the integrand. 
	
	Second, we turn to the three-particle cut in the $q^2$-channel, as presented in Figure~\ref{fig:AllCuts}$(iii)$. Importantly, the internal loop legs involve gluons with fixed helicity, rendering this cut completely universal and theory-independent. 
	
	Finally, we consider the three-particle cut in the $s_{23}$-channel, shown in Figure~\ref{fig:AllCuts}$(iv)$. In this case, the form factor entering the cut is non minimal and we have several possible helicity configurations for the momenta entering the loops, including fermions and scalars.
The relevant non-minimal form factors can be calculated with MHV diagrams \cite{Cachazo:2004kj} applied to form factors \cite{Dixon:2004za,Brandhuber:2011tv, Dawson:2014ora}
or more recent
methods introduced in \cite{Koster:2016loo, Chicherin:2016qsf,Koster:2016ebi}.
For convenience, we quote some of the non-minimal form factors entering the two-loop cut computations:%
	\footnote{See  also \cite{Dixon:1993xd,Neill:2009mz, Broedel:2012rc}  for related investigations of these quantities.}
\begin{align}
	\label{eq:treeFFs}
	\begin{split}
	&F_{\O_1}^{(0)}(1^+,2^+,3^+,4^-;q)\,=\, \frac{([12][23][31])^2}{[12][23][34][41]}\, , 
	\\[10pt]
	&F_{\O_1}^{(0)}(1^+,2^+,3^+,4^+;q)\,=\,\frac{[12][23][34][41]}{s_{12}}\left(1+\frac{[31][4|q|3\ra}{s_{23}[41]}\right)+\text{cyclic}(1,2,3,4)\,.
	\end{split}
\end{align}
Extracting  the integrand from the cut information is rather involved and we will present details of this calculation in  \cite{BKPTinprep}. 
With the help of the {\tt Mathematica} package {\tt LiteRed} \cite{Lee:2012cn,Lee:2013mka}
the two-loop integrand can be reduced to a basis of master integrals, whose explicit expressions 
were computed in  \cite{Gehrmann:1999as,Gehrmann:2000zt}. 
Finally, whenever possible we have simplified
the answer  by means of the symbol of transcendental functions \cite{Goncharov:2010jf}.

In the next section we use  the result of this calculation and present the  two-loop remainder function obtained after subtracting infrared divergences. 	
	
	
\section{Results}\label{sec:remainder}
\label{Sec:3}
	
	The two-loop remainder function of the form factor of a general operator $\O$ was first written  in \cite{Brandhuber:2012vm} using the same infrared subtraction scheme as its amplitude counterpart \cite{Anastasiou:2003kj,Bern:2005iz}. It is given by
\begin{align}\label{eq:remainder}
\cR^{(2)}_{\O} := \ \mathcal{F}_{\O}^{(2)}(\epsilon )\, - \, {1\over 2} \big( \mathcal{F}_{\O}^{(1)} (\epsilon) \big)^2 -  f^{(2)} (\epsilon)\  \mathcal{F}_{\O}^{(1)} ( 2 \epsilon ) 
+ \O (\epsilon )\,, 
\end{align}
where $\mathcal{F}^{(L)}_\mathcal{O}= F^{(L)}_\mathcal{O}/F^{(0)}_\mathcal{O}$,
$f^{(2)} (\eps) = -2(  \zeta_2 + \eps \, \zeta_3 + \eps^2 \, \zeta_4)$ and we have taken out a factor of $a [4 \pi e^{- \gamma_{\rm E}} (-\mu^2/q^2)]^{\epsilon}$ per loop. 

The remainder functions of scattering amplitudes or form factors of protected operators are finite quantities as they are free from ultraviolet  (UV) divergences. However, in the case of non-protected operators, the remainder does contain UV divergences. For the operator in question, 
we confirm that
all infrared  (IR) and mixed IR/UV  divergences cancel and all $1/{\epsilon^k}$
terms of the remainder vanish for $k=2,3,4$.

We find that the remainder contains a $1/\epsilon$ UV pole with coefficient $12-\pi^2 + \frac{1}{u v w}$.
The constant $-\pi^2$ arises from the subtraction scheme \eqref{eq:remainder} and is not part of the anomalous dimension, as in \cite{Brandhuber:2016fni}. The $\frac{1}{u v w}$ term is an indication of the mixing with the operator $\mathcal{O}_2$ 
introduced in 
Section \ref{Sec:2} (see \eqref{eq:operator2}).   Therefore we define
the one-loop corrected operator $\tilde{\mathcal{O}}_1 = \mathcal{O}_1 + C\,a\, \mathcal{O}_2$
and demand that the two-loop UV divergence of the form factor of $\tilde{\mathcal{O}}_1$ is proportional
to $F_{\O_1}^{(0)}(1^+,2^+,3^+;q)$ {\it i.e.}~the $\frac{1}{u v w}$ term is cancelled. This requirement fixes $C=1/6$ and the coefficient of the two-loop UV divergence to be $12$.
From this we infer the expected two-loop anomalous dimension of $\tilde\O_1$ as 
$\gamma_{\tilde\O_1}^{(2)}= -48\, a^2$, in agreement with that of the Konishi multiplet at this loop order. 
	
The finite part of the remainder of the form factor of $\O_1$ consists of functions of degree of transcendentality ranging from four to zero. We present it here in ``slices" of uniform transcendentality $m$, starting from the maximal degree four and denoting each slice as $\cR^{(2)}_{\O_1;m}$. The complete remainder is just the sum of all slices.
The answer is remarkably simple -- it contains only classical polylogarithms and, as we detail below, its building blocks are closely related to those of form factors of other non-protected operators \cite{Loebbert:2015ova,Brandhuber:2016fni,Loebbert:2016xkw}
hinting at a universal structure encompassing general classes of operators.

\newpage
	
\noindent{\bf Degree four:} The key observation is that the maximally transcendental part of the two-loop remainder of $\O_1$ is identical to that of the BPS operator $\O_{\rm BPS}=\Tr \, X^3$, computed in \cite{Brandhuber:2014ica} and already recognised as a universal building block in \cite{Brandhuber:2016fni,Loebbert:2015ova}:
\begin{align}\label{eq:remainderBPS}
	\cR^{(2)}_{\O_1;4}\ =\	\cR^{(2)}_{\rm BPS} \ = \ 
	& -\frac{3}{2}\, \text{Li}_4(u)+\frac{3}{4}\,\text{Li}_4\left(-\frac{u v}{w}\right) 
	-\frac{3}{2}\log(w) \, \text{Li}_3 \left(-\frac{u}{v} \right)+\frac{1}{16}  {\log}^2(u)\log^2(v) \nonumber \\
	&+{\log^2 (u) \over 32} \Big[ \log^2 (u) - 4 \log(v) \log(w) \Big]+{\zeta_2 \over 8 }\log(u) \Big[  5\log(u)- 2\log (v)\Big] \nonumber \\
	&+{\zeta_3 \over 2} \log(u) + \frac{7}{16}\, \zeta_4  + {\rm perms}\, (u,v,w) \, . 
\end{align}
	
\noindent{\bf Degree three:} At transcendentality three, new interesting structures appear as we get two types of terms: those consisting of pure transcendental functions and those multiplied by rational prefactors taken from the list $\left\{u/v,v/u, v/w,w/v, u/w, w/u\right\}$. 
The terms without any rational prefactors  take  the  form
	\begin{align}
	\cR^{(2)}_{{\O_1};3}\Big|_{\rm pure}\ =\ 
	&\text{Li}_3(u)+ \text{Li}_3(1-u)- {1\over 4} \log^2(u) \log \left({v w\over (1-u)^2} \right)
	+{1\over 3} \log (u) \log (v) \log (w)\nonumber \\
	&+\zeta_2 \log (u) -{5\over 3}\zeta_3  \, + \,\text{perms}\, (u,v,w)\,,
	\end{align}
which, remarkably,  is almost identical to the transcendentality-three part $\cR^{(2)}_{\text{non-BPS};3}$ of the two-loop remainder of the operator $\O_{\rm B}=\Tr(X[Y,Z])$ found in Eq.~$(4.11)$ of \cite{Brandhuber:2016fni}. Specifically,  we have (up to a $\log (-q^2)$ term)
\begin{align}
	\cR^{(2)}_{{\O_1};3}\Big|_{\rm pure} = {1\over 2}\Big(\cR^{(2)}_{\text{non-BPS};3}+
	4 \zeta_2\log(uvw)-24\, \zeta_3 \Big)\,.
\end{align}
We now move on to the terms with rational prefactors, which we label  by one of the possible
ratios listed above.
For concreteness we present the term with prefactor $u/w$:
\begin{align}
\label{tre}
	\cR^{(2)}_{{\O_1};3}\Big|_{u/w}=& \Big[-\text{Li}_3\left(-{u\over w}\right)+\log(u)\text{Li}_2\left({v\over 1-u}\right)-{1\over 2}\log(1-u)\log(u)\log\left({ w^2\over 1-u }\right)
	\nonumber \\
	&+{1\over 2}\text{Li}_3\left(-{uv\over w}\right)+{1\over 2}\log(u)\log(v)\log(w)+{1\over 12}\log^3(w) +(u\leftrightarrow v) \Big] \nonumber \\
	&+\text{Li}_3(1-v)-\text{Li}_3(u)+{1\over 2}\log^2(v)\log\left({1-v\over u}\right)\ 
	- \ 
	\zeta_2 \log\left( {u v\over w}\right) \,.
\end{align}
Another surprising observation can be made at this point.
Comparing \eqref{tre}  with the remainder density
$(R_i^{(2)})_{XXY}^{XYX}\Big|_3$ of form factors in the $SU(2)$ sector introduced in Eq.~$(3.22)$ of \cite{Loebbert:2015ova}, we observe that these are related (up to a $\log (-q^2)$ term), 
	\begin{align}
	\cR^{(2)}_{{\O_1};3}\Big|_{u/w}&=-(R_i^{(2)})_{XXY}^{XYX}\Big|_3\, - \, \zeta_2 \log \left({u} \right)\ . 
	\end{align}
The remaining terms, multiplied by different rational prefactors, follow the same pattern and can be simply found by taking the appropriate permutations of $u$, $v$ and $w$.
	
\noindent{\bf Degree two:} At transcendentality two, again we have two types of terms -- those consisting of purely transcendental functions and those multiplied by rational coefficients. The pure part reads
	\begin{align}
	\cR^{(2)}_{{\O_1};2}\Big|_{\rm pure}&= -\text{Li}_2(1-u)-\log^2(u)+{1\over 2}\log(u)\log(v)-{13\over2}\zeta_2 \, + \,\text{perms}\, (u,v,w)\,.
	\end{align}
	The other part consists of terms multiplied by one of the following rational coefficients: $\left\{u^2/v^2, v^2/ u^2, u^2/ w^2, v^2/ w^2,w^2/ u^2, w^2/ v^2\right\}$. The  term multiplied by $u^2/w^2$ has the form: 
	\begin{align}
	\cR^{(2)}_{{\O_1};2}\Big|_{u^2/w^2}&=  \text{Li}_2(1 - u) + \text{Li}_2(1 - v) + \log(u) \log(v)-\zeta_2\,,
	\end{align}
where again the remaining terms are obtained by appropriate permutations of $u$, $v$ and $w$.
	
\noindent{\bf Degree one and zero:} The degree-one terms can be presented in a very compact form as
\begin{align}
\cR^{(2)}_{{\O_1};1}&= \left(-4 + \frac{v }{w}+ \frac{u^2 }{2 v w}  \right) \log(u)\, + \,\text{perms}\, (u,v,w)\, ,
\end{align}
while the degree-zero terms read
\begin{align}\label{smile}
\cR^{(2)}_{{\O_1};0}&=7  \left(12+\frac{1}{uvw} \right)\,.
\end{align}
As a final comment we note that the constant part of \eqref{smile}
times $-4/7$ equals the value of the two-loop Konishi anomalous dimension -- the
same observation  was made in \cite{Loebbert:2015ova} for operators in the $SU(2)$ sector.
	
\section{Beyond $\N\!=\!4$ SYM}
\label{Sec:4}


In this final section, we argue that the universality of the maximally transcendental part of the remainder function of $F_{\O_1}(1^+,2^+,3^+;q)$ is not confined to $\N\!=\!4$ SYM, and in fact extends to theories with less supersymmetry, including pure Yang-Mills and QCD.

All deviations from $\N\!=\!4$ SYM are due to a different matter content (scalars and fermions), and we now analyse how these  affect the cuts in Figure \ref{fig:AllCuts}. First, we note that the diagrams $(i)$ and $(iii)$ are purely gluonic and, therefore, theory-independent. Second, the diagram in Figure \ref{fig:AllCuts}$(ii)$ contains a four-point one-loop amplitude. If the matter content is changed compared to $\cN\!=\!4$ SYM, this amplitude will receive modifications through additional bubble integrals \cite{Bern:1994cg, Bern:1993mq,Bedford:2004nh}, which can only produce two-loop integrals of lower transcendentality.
This leaves us with  Figure \ref{fig:AllCuts}$(iv)$, and we need to analyse the individual contributions from fermions and scalars propagating across the cut.
\begin{figure}[h]
		\centering
		\includegraphics[width=0.3\linewidth]{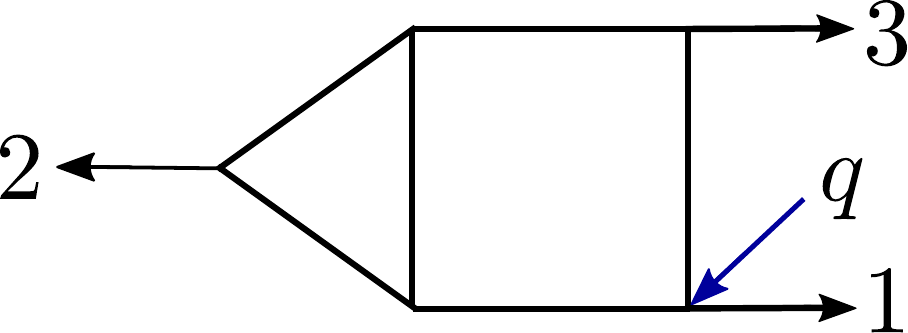}
		\caption{\it The single-scale integral topology which incorporates the effect of having a different field content compared to that of $\N\!=\!4$ SYM.}
		\label{fig:SOTB}
	\end{figure}\\
Our  computation shows that  such  contributions appear through  the integral topology shown in Figure \ref{fig:SOTB} which, due to non-trivial cancellations,  is absent for $\N\!=\!4$ SYM. Evaluating explicitly the integrals with appropriate numerators coming from fermions and scalars crossing the cut, we find again that they only contribute at sub-maximal transcendental weight.
	Hence we conclude that the transcendentality-four slice of the remainder function is indeed universal for this particular form factor in Yang-Mills theories with any amount of supersymmetry and QCD (the presence of fermions in the fundamental representation does not alter this statement).
	
	
We end by commenting on possible extensions of our work that are  currently under
investigation \cite{BKPTinprep}. An obvious important step is to generalise our calculation to theories with
less supersymmetry, including pure Yang-Mills and QCD. Here it will be important to address
potential rational terms that may be missed in less supersymmetric theories when four-dimensional cuts are employed (rather than $D$-dimensional ones). Note that issues encountered with dimensional regularisation in the case of the Konishi operator in \cite{Loebbert:2015ova}
did not arise in \cite{Brandhuber:2016fni} and in  the present work since the operator definition does not involve state sums. Other aspects to be discussed in future work are form factors of
other dimension-six operators such as ${\rm Tr}(DFDF)$ appearing in the effective theory for
Higgs plus multi-parton scattering, and studies of the operator mixing using subminimal/non-minimal form factors as in \cite{Brandhuber:2016fni}. Finally, we are also investigating form factors
with more general helicity configurations than the one considered in this letter.
We expect that in all these considerations supersymmetry will emerge as a powerful organisational principle and that results for form factors in QCD will reveal further remarkable similarities with $\N\!=\!4$ SYM.

\section*{Acknowledgements}
	
We would like to thank Zvi Bern, John Joseph Carrasco  and Henrik Johansson for very helpful discussions, Claude Duhr for sharing a package for handling polylogarithms, and Massimo Bianchi, Sophia Borowka, Claude Duhr, Sergio Ferrara, Paul Heslop, Jan Plefka, Emery Sokatchev,  Yassen Stanev, Massimo Testa and Donovan Young for stimulating conversations. 
The work of AB and GT was supported by the Science and Technology Facilities Council (STFC) Consolidated Grant ST/L000415/1  
\textit{``String theory, gauge theory \& duality"}. The work of MK is supported by an STFC quota studentship. BP is funded by the ERC starting grant 637019 ``\emph{MathAm}''. AB and GT would like to thank the KITP at the University of California, Santa Barbara, where their research was supported by the National Science Foundation under Grant No.~NSF PHY-1125915.
GT is grateful to the Alexander von Humboldt Foundation for support through a Friedrich Wilhelm Bessel Research Award, and to the Institute for Physics and IRIS Adlershof at Humboldt University, Berlin, for their warm hospitality.

	\bibliographystyle{utphys}
	\bibliography{remainder}

\end{document}